\documentclass{appolb}
\usepackage{graphicx}
\usepackage{amsmath}
\usepackage[citestyle=ieee,
natbib=true,
style=numeric,
giveninits=true,
backend=biber,
date=year]{biblatex}
\usepackage{macros}
\usepackage{authblk}
\usepackage{color}
\usepackage{ulem}

\DeclareFieldFormat[article]{title}{}

\begin{document}
\title{Maximum mass and stability of differentially rotating neutrons stars%
\thanks{Presented at The 8th Conference of the Polish Society on Relativity}%
}

\author{Paweł Szewczyk, Dorota Gondek-Rosińska
\address{Astronomical Observatory Warsaw University, 00-478 Warsaw, Poland}
\\
{Pablo Cerd\'a-Dur\'an
\address{ Observatori Astronòmic, Universitat de València, E-46980, Paterna (València), Spain}
}
}

\maketitle
\begin{abstract}
We present our study of stability of differentially rotating, axisymmetric neutron stars described by a polytropic equation of state with $\Gamma = 2$.
We focus on quasi-toroidal solutions with a degree of differential rotation $\widetilde A=1$.
Our results show that for a wide range of parameters hypermassive, quasi-toroidal neutron stars are dynamically stable against quasi-radial perturbations, which may have implications for newly born neutron stars and binary neutron stars mergers.
\end{abstract} 

\section{Introduction}
Differential rotation seems to appear naturally in many dynamical scenarios involving neutron stars (NS), including the collapse of stellar cores (see e.g. \citep{Villain2004}) and binary neutron star (BNS) mergers (see e.g. \citep{Kastaun2015}). Its stabilizing effect may allow for configurations with masses significantly higher than the mass limit for rigidly rotating neutron stars. Its study is relevant for the understanding of black hole formation in those astrophysical scenarios with consequences in observations of core-collapse supernovae (CCSN) and BNS mergers, especially with current gravitational wave ground-based observatories (LIGO, Virgo and Kagra \cite{LIGO:2015, Virgo:2015, KAGRA:2019}) and future ones (the Einstein Telescope and Cosmic Explorer).

\subsection{Equilibrium models of differentially rotating NS}

The solution space of differentially rotating neutron stars in equilibrium was already extensively studied by different authors.
It was shown by \citep{Baumgarte2000} that differentially rotating NS with masses significantly larger than non-rotating or rigidly rotating NS can exist and be stable against radial collapse and bar formation. Those with masses larger than the limit for rigidly rotating objects are called hypermassive NS.

However, studying the whole solution space has proven to be numerically challenging.
The existence of different types of solutions of differentially rotating neutron stars was for the first time found by \citep{Ansorg2009}  using relativistic highly accurate and stable multi-domain spectral numerical code FlatStar.
Most importantly, for a given degree of differential rotation, the solution is not uniquely determined by the maximal density and angular momentum of the NS (or any other suitable pair of parameters), as is the case for rigid rotation. Instead,
different types of solutions may coexist for the same parameters.
The maximum mass for different degrees of differential rotation and different solution types was presented by \citep{Rosinska2017} and \citep{Studzinska2016} for polytropes, showing that the most massive configurations are obtained for modest degree of differential rotation. 
Similar results were obtained for strange quark stars \citep{Szkudlarek2019} and NS with several realistic equations of state \citep{Espino2019}.

While many studies in the past were using a rotation law of \citet{komatsu89II}, which is mainly consistent with CCSN remnants \cite{Villain2004}, the rotation law observed 
in simulations of BNS merger remnants departs significantly from that one. Rotation laws better suited for BNS mergers have been proposed by \citep{Uryu2017}, and its impact on the solution space of equilibrium models studied by \citep{Iosif21}.

\subsection{Stability properties of hypermassive NS}

For non-rotating NS the limit for both secular and dynamical stability occurs at the point of maximal mass ($M_{TOV}$).
This criterion can be, to some point, extended to rigidly rotating NS.
The so-called turning point criterion was presented by \citet{Friedman88} and proven to be a sufficient criterion for instability. 
It states that the point of maximal gravitational mass $M$ on a sequence of configurations of fixed angular momentum $J$ ($J$-constant turning points), or, alternatively, the point of minimal gravitational mass on a sequence of fixed rest mass $M_0$ ($M_0$-constant turning points) marks the onset of instability.
This criterion, however, does not give the exact threshold to collapse.
The neutral-stability point where F-mode frequency vanishes differs from the turning-point line \citep{Takami2011}.
Numerical simulations confirm that the neutral-stability line marks the threshold to prompt collapse.

For rigidly rotating NS, the $J$-constant turning points coincide with the $M_0$-constant turning points, but it is no longer the case for differential rotation of a given degree. While other authors usually refer to the former, in this paper we use the latter as we find it to be a closer estimate of stability threshold.

On secular timescales, differential rotation transforms into rigid rotation due to effects of viscosity and magnetic breaking \citep{Shapiro2000, Duez2004, Shibata2005}.
By definition, hypermassive NS have masses that cannot be supported by rigid rotation only.
This eventually may lead to a delayed collapse and delayed emission of gravitational waves.
There is no clear criterion of dynamical stability for hypermassive NS to tell if the collapse will be prompt or delayed.

Various authors have studied the stability properties of differentially rotating NS by means of numerical simulations.
An example of hypermassive NS dynamically stable to both radial instabilities and bar formation was presented by \cite{Baumgarte2000}. In \citep{Weih2018} the authors explore the limit of stability to quasi-radial oscillations for differentially rotating NS, excluding quasi-toroidal configurations. The threshold to collapse proves to be close to the ($J$-constant) turning-point line, which is still a valid sufficient criterion of dynamical instability. A caveat for the large masses supported by many these works is that they may be subject to non-axisymmetric corrotational instabilities (usually known as low-$T/|W|$ instabilities, see e.g. \cite{Shibata2003}) that are able to transport efficiently angular momentum and erase differential rotation. Although there is no clear criterion for the onset of this instability, all studied cases in the literature of NS with quasi-toroidal shape (e.g. \cite{Espino2019-stab}) have shown the dynamical growth of these instabilities.

\section{Equilibrium models}

We consider axisymmetric, stationary configuration of rotating fluid in cylindrical coordinates $(t, \rho, z, \phi)$.
The configurations we study are highly flattened and cylindrical coordinates are more practical than spherical ones.
The line element associated with such configuration may be written in form:

\begin{equation}
\label{eq:grmetric}
    ds^2 = - e^{2\nu} dt^2 + e^{2\mu}(d\rho^2 + dz^2) + W^2 e^{-2\nu} (d\phi - \omega dt)^2 \text{,}
\end{equation}

with four metric potentials $\mu, W, \nu, \omega$, being functions of $\rho$ and $z$ only, due to symmetry.

In general, the properties of matter are defined by the equation of state (EOS).
Here we use a polytropic EOS with $\Gamma = 2$ (or $N = 1$ in alternate notation) which yields a relation between total mass-energy density $\epsilon$ and pressure $p$:
\begin{equation}
    \epsilon(p) = p + \sqrt{\frac{p}{K}} \text{,}
\end{equation}

where $K$ is the polytropic constant.

We often use a dimensionless value of relativistic enthalpy as a main thermodynamical parameter, which in case of polytrope with $\Gamma = 2$ can be expressed as:

\begin{equation}
    H = \text{log}(1 + 2K\epsilon_B)\text{,}
\end{equation}

where $\epsilon_B$ is the rest-mass density.

To describe the differential rotation profile, one need to specify the rotation law. For the fluid four-velocity $u^\alpha$ and angular velocity $\Omega$, it can be defined as function $u^t u_\phi = F(\Omega)$.
Here we use the J-const rotation law \citep{Friedman88}:

\begin{equation}
    F(\Omega) = A^2(\Omega_c - \Omega) \text{,}
\end{equation}

with $\Omega_c$ being the angular velocity on the rotation axis and $A$ a constant describing the steepness of the profile.
This produces a rotation profile consistent with remnants of CCSN.
As a dimensionless measure of the degree of differential rotation, we use the value of $\widetilde A = \frac{r_e}{A}$ with $r_e$ being the star radius on equatorial plane.

To construct the initial data we solve the relativistic field equations for four metric potentials on the $\rho,z$ plane using the highly accurate code FlatStar.
It uses an efficient multi-domain spectral method to construct equilibrium models of rotating compact objects.
For more technical details see \citep{Ansorg2009} and appendix of \citep{Rosinska2017}.

\begin{figure}[htb]
    \centering
    \includegraphics[width=12.5cm]{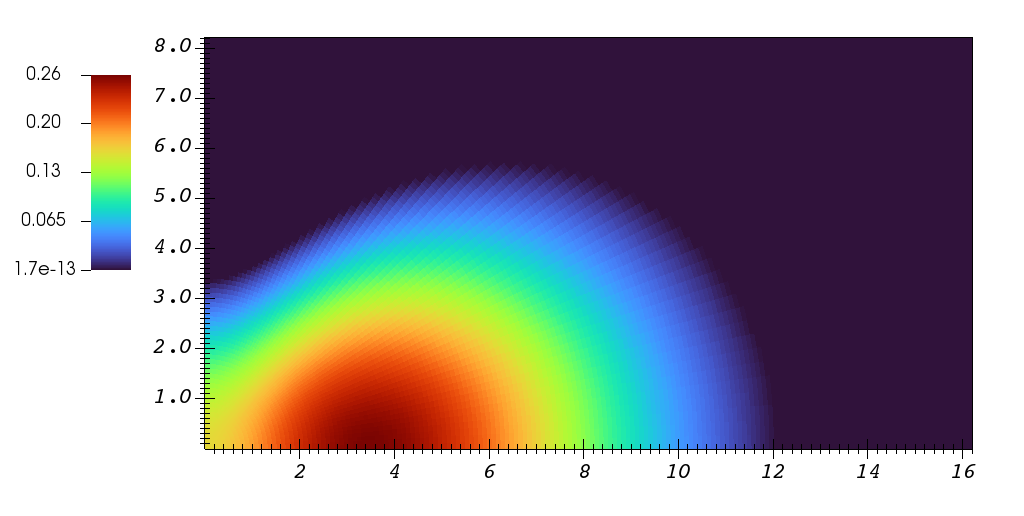}
    \caption{Example of a quasi-toroidal initial configuration ($H_{\max} = 0.26$, $M_0 = 0.33$). Color coded, the relativistic enthalpy $H$ in a meridional cross section. The maximal value $H_{\max}$ is found far off the rotation axis ($y=0$). }
    \label{fig:profile}
\end{figure}

In this paper, we focus on the degree of differential rotation $\widetilde A = 1$.
According to classification of \citep{Ansorg2009}, all configurations studied here are of type C.
Our main interest lies in quasi-toroidal configurations, which produce the largest masses and are not extensively studied by other authors.
Figure \ref{fig:profile} shows the relativistic enthalpy profile of one of such configurations.
The maximal value, $H_{\max}$, is not in the center of star, and hence the classification as quasi-toroidal. We select 5 sequences with constant rest mass of configurations close to the stability limit estimated by the turning point criterion.

\section{Stability against quasi-radial instabilities}

To test the stability of the selected configurations, we perform relativistic axisymmetric hydrodynamical simulations using the CoCoNuT code \citep{cerda-duran2008}, which uses the conformally flat approximation (CFC). In the 3+1 split, the line element reads:
\begin{equation}
\label{ADM}
    ds^2 = -\alpha^2 dt^2 + \gamma_{ij}(dx^i + \beta^i dt)(dx^j + \beta^j dt),
\end{equation}
where $\gamma_{ij} = \Phi^4 \delta_{ij}$ in the CFC approximation, with the conformal factor $\Phi^6 =e^{2\mu - u}$ (using variables from equation \ref{eq:grmetric}).
The accuracy of CFC was tested, for example, by \citep{cst96}, showing that the astrophysical properties (such as mass) are reconstructed with only a small discrepancy less than 5\%.

To induce the collapse in unstable configurations, we introduce a radial perturbation in form of an additional velocity component.
The amplitude of this perturbation is carefully chosen to make the mean value of the maximal density match the initial value in stable solutions (for unstable solutions we extrapolate the amplitude from the stable region). We test our results by comparing them with the results of \citep{Weih2018} and \citep{Takami2011}, finding them to be in agreement.

\begin{figure}[htb]
    \centering
    \includegraphics[width=12.5cm]{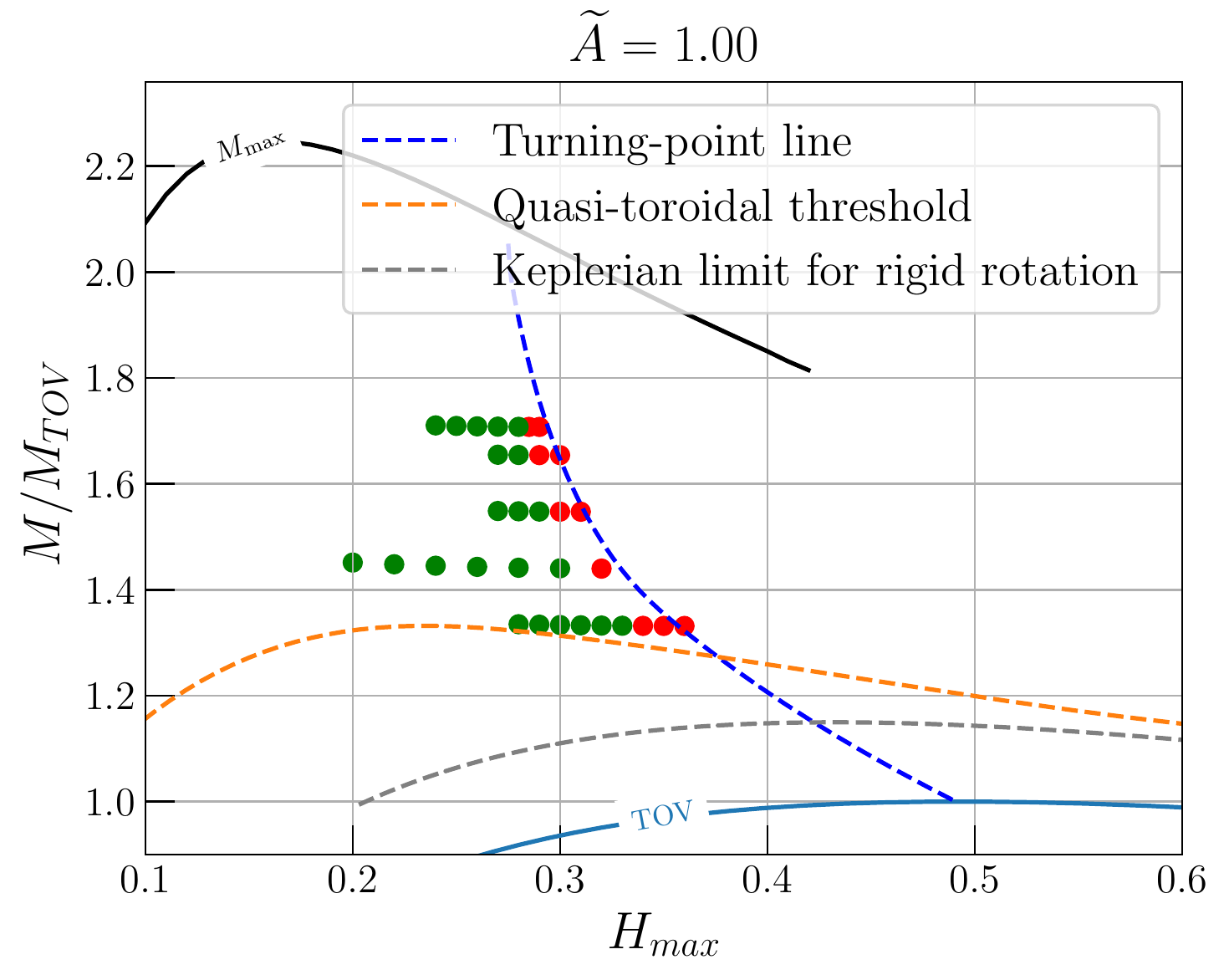}
    \caption{Simulated configurations divided into stable (green marks) and unstable (red marks) to quasi-radial perturbations. Blue dashed line shows the line of ($M_0$-constant) turning-points, being the first estimate of stability. The orange dashed line marks the boundary between spheroidal and quasi-toroidal configurations. Limit of mass for this degree of differential rotation, limit for rigid rotation and sequence of non-rotating NS are presented for reference.}
    \label{fig:c10}
\end{figure}

For all our models we inspect the evolution of maximal density in time.
For stable configurations we see the value of density oscillating around the initial value.
Unstable configurations show an exponential growth of the maximal density in the first few ms, marking the prompt collapse to a BH.
Figure \ref{fig:c10} shows stable and unstable configurations on $H_{\max}-M$ plane.

\section{Summary}

We have selected a sample of quasi-toroidal configurations of neutron stars with polytropic equation of state.
We used a polytrope with $\Gamma=2$ and j-constant rotation law.
By performing a numerical relativistic hydrodynamical evolution, we tested stability of the selected equilibria against axisymmetrical (quasi-radial) perturbations.
We show that differential rotation allows the existence of dynamically stable models with masses almost twice as massive as $M_{\rm TOV}$.
These stable configurations, if formed during CCSN or BNS mergers, may undergo a significantly delayed collapse to a black hole.
Further study is needed to inspect the stability properties against non-axisymmetrical perturbations on dynamical timescales.

\section{Acknowledgments}

This work was partially supported by the Polish National Science Centre grants No. 2017/26/M/ST9/00978 and 2022/45/N/ST9/04115, by POMOST/2012-6/11
Program of Foundation for Polish Science co-financed
by the European Union within the European Regional
Development Fund, by the Spanish Agencia Estatal de Investigaci\'on (Grants No. PGC2018-095984-B-I00 and PID2021-125485NB-C21) funded by MCIN/AEI/10.13039/501100011033 and ERDF A way of making Europe, by the Generalitat Valenciana (PROMETEO/2019/071), and by COST Actions CA16104 and CA16214.

\printbibliography

@ARTICLE{Shibata2005,
   author = {{Shibata}, M. and {Taniguchi}, K. and {Ury{\= u}}, K.},
    title = "{Merger of binary neutron stars with realistic equations of state in full general relativity}",
  journal = {\prd},
   eprint = {gr-qc/0503119},
     year = 2005,
   volume = 71,
   number = 8,
      eid = {084021},
    pages = {084021},
      doi = {10.1103/PhysRevD.71.084021},
   adsurl = {http://adsabs.harvard.edu/abs/2005PhRvD..71h4021S}
}

@ARTICLE{Rosinska2017,
       author = {{Gondek-Rosi{\'n}ska}, D. and {Kowalska}, I. and {Villain}, L. and {Ansorg}, M. and {Kucaba}, M.},
        title = "{A New View on the Maximum Mass of Differentially Rotating Neutron Stars}",
      journal = {\apj},
         year = "2017",
       volume = {837},
          eid = {58},
        pages = {58},
          doi = {10.3847/1538-4357/aa56c1},
archivePrefix = {arXiv},
       eprint = {1609.02336},
 primaryClass = {astro-ph.HE},
       adsurl = {https://ui.adsabs.harvard.edu/\#abs/2017ApJ...837...58G}
}

@article{Ansorg2009,
    author = {{Ansorg}, M. and {Gondek-Rosińska}, D. and {Villain}, L.},
    year = {2008},
    pages = {},
    title = {On the Solution Space of Differentially Rotating Neutron Stars in General Relativity},
    volume = {396},
    journal = {\mnras},
    doi = {10.1111/j.1365-2966.2009.14904.x}
}

@ARTICLE{komatsu89II,
   author = {{Komatsu}, H. and {Eriguchi}, Y. and {Hachisu}, I.},
    title = "{Rapidly rotating general relativistic stars. II - Differentially rotating polytropes}",
  journal = {\mnras},
     year = 1989,
   volume = 239,
    pages = {153-171},
      doi = {10.1093/mnras/239.1.153},
   adsurl = {http://adsabs.harvard.edu/abs/1989MNRAS.239..153K}
}

@ARTICLE{Friedman88,
   author = {{Friedman}, J.~L. and {Ipser}, J.~R. and {Sorkin}, R.~D.},
    title = "{Turning-point method for axisymmetric stability of rotating relativistic stars}",
  journal = {\apj},
     year = 1988,
   volume = 325,
    pages = {722-724},
      doi = {10.1086/166043},
   adsurl = {https://ui.adsabs.harvard.edu/abs/1988ApJ...325..722F}
}

@article{Takami2011,
    author = {Takami, K. and Rezzolla, L. and Yoshida, S.},
    title = "{A quasi-radial stability criterion for rotating relativistic stars}",
    journal = {MNRAS: Letters},
    volume = {416},
    number = {1},
    pages = {L1-L5},
    year = {2011},
    issn = {1745-3925},
    doi = {10.1111/j.1745-3933.2011.01085.x},
    url = {https://doi.org/10.1111/j.1745-3933.2011.01085.x},
    eprint = {http://oup.prod.sis.lan/mnrasl/article-pdf/416/1/L1/6140636/416-1-L1.pdf},
}

@ARTICLE{Weih2018,
       author = {{Weih}, L. R. and {Most}, E. R. and {Rezzolla}, L.},
        title = "{On the stability and maximum mass of differentially rotating relativistic stars}",
      journal = {\mnras},
         year = "2018",
       volume = {473},
       number = {1},
        pages = {L126-L130},
          doi = {10.1093/mnrasl/slx178},
archivePrefix = {arXiv},
       eprint = {1709.06058},
 primaryClass = {gr-qc},
       adsurl = {https://ui.adsabs.harvard.edu/abs/2018MNRAS.473L.126W}
}

@ARTICLE{Espino2019-stab,
       author = {{Espino}, Pedro L. and {Paschalidis}, Vasileios and {Baumgarte}, Thomas W. and {Shapiro}, Stuart L.},
        title = "{Dynamical stability of quasitoroidal differentially rotating neutron stars}",
      journal = {\prd},
         year = 2019,
        month = aug,
       volume = {100},
       number = {4},
          eid = {043014},
        pages = {043014},
          doi = {10.1103/PhysRevD.100.043014},
archivePrefix = {arXiv},
       eprint = {1906.08786},
 primaryClass = {astro-ph.HE},
       adsurl = {https://ui.adsabs.harvard.edu/abs/2019PhRvD.100d3014E}
}

@ARTICLE{Espino2019,
       author = {{Espino}, P. L. and {Paschalidis}, V.},
        title = "{Revisiting the maximum mass of differentially rotating neutron stars in general relativity with realistic equations of state}",
      journal = {\prd},
         year = "2019",
       volume = {99},
       number = {8},
          eid = {083017},
        pages = {083017},
          doi = {10.1103/PhysRevD.99.083017},
archivePrefix = {arXiv},
       eprint = {1901.05479},
 primaryClass = {astro-ph.HE},
       adsurl = {https://ui.adsabs.harvard.edu/abs/2019PhRvD..99h3017E}
}

@ARTICLE{Studzinska2016,
   author = {{Studzi{\'n}ska}, A.~M. and {Kucaba}, M. and {Gondek-Rosi{\'n}ska}, D. and {Villain}, L. and {Ansorg}, M.},
    title = "{Effect of the equation of state on the maximum mass of differentially rotating neutron stars}",
  journal = {\mnras},
     year = 2016,
   volume = 463,
    pages = {2667-2679},
      doi = {10.1093/mnras/stw2152},
   adsurl = {https://ui.adsabs.harvard.edu/abs/2016MNRAS.463.2667S}
}

@ARTICLE{Szkudlarek2019,
   author = {{Szkudlarek}, M. and {Gondek-Rosi{\'n}ska}, D. and {Villain}, L. and {Ansorg}, M.},
    title = "{Maximum Mass of Differentially Rotating Strange Quark Stars}",
  journal = {\apj},
archivePrefix = "arXiv",
   eprint = {1904.03759},
 primaryClass = "astro-ph.HE",
     year = 2019,
    month = jul,
   volume = 879,
      eid = {44},
    pages = {44},
      doi = {10.3847/1538-4357/ab1752},
   adsurl = {https://ui.adsabs.harvard.edu/abs/2019ApJ...879...44S}
}

@ARTICLE{Baumgarte2000,
       author = {{Baumgarte}, Thomas W. and {Shapiro}, Stuart L. and {Shibata}, Masaru},
        title = "{On the Maximum Mass of Differentially Rotating Neutron Stars}",
      journal = {\apjl},
         year = "2000",
        month = "Jan",
       volume = {528},
       number = {1},
        pages = {L29-L32},
          doi = {10.1086/312425},
archivePrefix = {arXiv},
       eprint = {astro-ph/9910565},
 primaryClass = {astro-ph},
       adsurl = {https://ui.adsabs.harvard.edu/abs/2000ApJ...528L..29B}
}

@ARTICLE{cerda-duran2008,
       author = {{Cerd{\'a}-Dur{\'a}n}, P. and {Font}, J.~A. and {Ant{\'o}n}, L. and {M{\"u}ller}, E.},
        title = "{A new general relativistic magnetohydrodynamics code for dynamical spacetimes}",
      journal = {\aap},
         year = 2008,
        month = dec,
       volume = {492},
       number = {3},
        pages = {937-953},
          doi = {10.1051/0004-6361:200810086},
archivePrefix = {arXiv},
       eprint = {0804.4572},
 primaryClass = {astro-ph},
       adsurl = {https://ui.adsabs.harvard.edu/abs/2008A&A...492..937C}
}

@ARTICLE{Iosif21,
       author = {{Iosif}, Panagiotis and {Stergioulas}, Nikolaos},
        title = "{Equilibrium sequences of differentially rotating stars with post-merger-like rotational profiles}",
      journal = {\mnras},
         year = 2021,
        month = may,
       volume = {503},
       number = {1},
        pages = {850-866},
          doi = {10.1093/mnras/stab392},
archivePrefix = {arXiv},
       eprint = {2011.10612},
 primaryClass = {gr-qc},
       adsurl = {https://ui.adsabs.harvard.edu/abs/2021MNRAS.503..850I}
}

@article{Uryu2017,
  title = {Modeling differential rotations of compact stars in equilibriums},
  author = {Uryu, K. and Tsokaros, Antonios and Baiotti, Luca and Galeazzi, Filippo and Taniguchi, Keisuke and Yoshida, Shin'ichirou},
  journal = {\prd},
  volume = {96},
  issue = {10},
  pages = {103011},
  numpages = {8},
  year = {2017},
  month = {Nov},
  publisher = {American Physical Society},
  doi = {10.1103/PhysRevD.96.103011},
  url = {https://link.aps.org/doi/10.1103/PhysRevD.96.103011}
}

@ARTICLE{Shapiro2000,
       author = {{Shapiro}, Stuart L.},
        title = "{Differential Rotation in Neutron Stars: Magnetic Braking and Viscous Damping}",
      journal = {\apj},
         year = 2000,
        month = nov,
       volume = {544},
       number = {1},
        pages = {397-408},
          doi = {10.1086/317209},
archivePrefix = {arXiv},
       eprint = {astro-ph/0010493},
 primaryClass = {astro-ph},
       adsurl = {https://ui.adsabs.harvard.edu/abs/2000ApJ...544..397S}
}

@article{Kastaun2015,
  title = {Properties of hypermassive neutron stars formed in mergers of spinning binaries},
  author = {Kastaun, Wolfgang and Galeazzi, Filippo},
  journal = {\prd},
  volume = {91},
  issue = {6},
  pages = {064027},
  numpages = {17},
  year = {2015},
  month = {Mar},
  publisher = {American Physical Society},
  doi = {10.1103/PhysRevD.91.064027},
  url = {https://link.aps.org/doi/10.1103/PhysRevD.91.064027}
}

@ARTICLE{Villain2004,
       author = {{Villain}, L. and {Pons}, J.~A. and {Cerd{\'a}-Dur{\'a}n}, P. and {Gourgoulhon}, E.},
        title = "{Evolutionary sequences of rotating protoneutron stars}",
      journal = {\aap},
         year = 2004,
        month = apr,
       volume = {418},
        pages = {283-294},
          doi = {10.1051/0004-6361:20035619},
archivePrefix = {arXiv},
       eprint = {astro-ph/0310875},
 primaryClass = {astro-ph},
       adsurl = {https://ui.adsabs.harvard.edu/abs/2004A&A...418..283V}
}

@ARTICLE{Duez2004,
       author = {{Duez}, Matthew D. and {Liu}, Yuk Tung and {Shapiro}, Stuart L. and {Stephens}, Branson C.},
        title = "{General relativistic hydrodynamics with viscosity: Contraction, catastrophic collapse, and disk formation in hypermassive neutron stars}",
      journal = {\prd},
         year = 2004,
        month = may,
       volume = {69},
       number = {10},
          eid = {104030},
        pages = {104030},
          doi = {10.1103/PhysRevD.69.104030},
archivePrefix = {arXiv},
       eprint = {astro-ph/0402502},
 primaryClass = {astro-ph},
       adsurl = {https://ui.adsabs.harvard.edu/abs/2004PhRvD..69j4030D}
}

@ARTICLE{cst96,
       author = {{Cook}, Gregory B. and {Shapiro}, Stuart L. and {Teukolsky}, Saul A.},
        title = "{Testing a simplified version of Einstein's equations for numerical relativity}",
      journal = {\prd},
         year = 1996,
        month = may,
       volume = {53},
       number = {10},
        pages = {5533-5540},
          doi = {10.1103/PhysRevD.53.5533},
archivePrefix = {arXiv},
       eprint = {gr-qc/9512009},
 primaryClass = {gr-qc},
       adsurl = {https://ui.adsabs.harvard.edu/abs/1996PhRvD..53.5533C}
}

@ARTICLE{Virgo:2015,
       author = {{Acernese}, F. and {Agathos}, M. and others},
        title = "{Advanced Virgo: a second-generation interferometric gravitational wave detector}",
      journal = {Classical and Quantum Gravity},
         year = 2015,
        month = jan,
       volume = {32},
       number = {2},
          eid = {024001},
        pages = {024001},
          doi = {10.1088/0264-9381/32/2/024001},
archivePrefix = {arXiv},
       eprint = {1408.3978},
 primaryClass = {gr-qc},
       adsurl = {https://ui.adsabs.harvard.edu/abs/2015CQGra..32b4001A}
}

@ARTICLE{LIGO:2015,
       author = {{LIGO Scientific Collaboration} and {Aasi}, J. and {Abbott}, B.~P. and others},
        title = "{Advanced LIGO}",
      journal = {Classical and Quantum Gravity},
         year = 2015,
        month = apr,
       volume = {32},
       number = {7},
          eid = {074001},
        pages = {074001},
          doi = {10.1088/0264-9381/32/7/074001},
archivePrefix = {arXiv},
       eprint = {1411.4547},
 primaryClass = {gr-qc},
       adsurl = {https://ui.adsabs.harvard.edu/abs/2015CQGra..32g4001L}
}

@ARTICLE{KAGRA:2019,
       author = {{Kagra Collaboration} and {Akutsu}, T. and {Ando}, M. and others},
        title = "{KAGRA: 2.5 generation interferometric gravitational wave detector}",
      journal = {Nature Astronomy},
         year = 2019,
        month = jan,
       volume = {3},
        pages = {35-40},
          doi = {10.1038/s41550-018-0658-y},
archivePrefix = {arXiv},
       eprint = {1811.08079},
 primaryClass = {gr-qc},
       adsurl = {https://ui.adsabs.harvard.edu/abs/2019NatAs...3...35K}
}

@ARTICLE{Shibata2003,
       author = {{Shibata}, Masaru and {Karino}, Shigeyuki and {Eriguchi}, Yoshiharu},
        title = "{Dynamical bar-mode instability of differentially rotating stars: effects of equations of state and velocity profiles}",
      journal = {\mnras},
         year = 2003,
        month = aug,
       volume = {343},
       number = {2},
        pages = {619-626},
          doi = {10.1046/j.1365-8711.2003.06699.x},
archivePrefix = {arXiv},
       eprint = {astro-ph/0304298},
 primaryClass = {astro-ph},
       adsurl = {https://ui.adsabs.harvard.edu/abs/2003MNRAS.343..619S}
}

\end{document}